\documentstyle[12pt]{article}
\include{epsf}
\begin{document}
\centerline{\Large \bf Effect of Screening of Intermicellar Interactions on the }
\vspace{0.25cm}
\centerline{\Large \bf Linear and Nonlinear Rheology of a Viscoelastic Gel}
\vspace{0.5cm}
\centerline{\bf Ranjini Bandyopadhyay and A. K. Sood}
\vspace{0.5cm}
\centerline{\it Department of Physics, Indian Institute of Science, Bangalore 560 012, India}
\vspace{1cm}
\centerline{\Large \bf Abstract}
\vspace{0.5cm}

\noindent We report our studies of the linear and nonlinear rheology of aqueous solutions of 
the surfactant cetyl trimethylammonium tosylate (CTAT) with varying amounts of sodium 
chloride (NaCl). The CTAT concentration is fixed at 42mM and the salt concentration is 
varied between 0mM to 120mM. On increasing the salt (NaCl) concentration, we see three 
distinct regimes in the zero-shear viscosity and the high frequency plateau modulus data. 
In regime I, the zero-shear viscosity shows a weak increase with salt concentration due 
to enhanced micellar growth. The decrease in the zero-shear viscosities with salt 
concentration in regimes II and III can be explained in terms of inter-micellar branching. 
The most intriguing feature of our data  however, is the anomalous behavior of the high frequency plateau modulus in regime II (0.12 $\le \frac{[NaCl]}{[CTAT]} \le$ 1.42). In 
this regime, the plateau modulus {\it increases} with an increase in NaCl concentration. 
This is highly counter-intuitive, since the correlation length of concentration fluctuations 
and hence the plateau modulus $G_{\circ}$ are not expected to change appreciably in the 
semi-dilute regime. We propose to explain the changes in regime II in terms of the unbinding 
of the organic counterions (tosylate) from the CTA$^{+}$ surfaces on the addition of NaCl. 
In the nonlinear flow curves of the samples with high salt content, significant deviations 
from the predictions of the Giesekus model for entangled micelles are observed. 
\newpage
\section{Introduction}

The flow behavior of surfactant solutions have been studied extensively, both theoretically 
and experimentally \cite{rehage,ch5cates}. Viscoelastic gels, which are formed by 
the entanglement of cylindrical micelles under appropriate conditions of temperature 
and salinity and in the presence of suitable counterions, can be used as model systems 
for rheological research. Even though the internal structures of these materials are often 
very complex, the viscoelastic parameters characterizing them are found to follow very 
simple scaling laws,  making the prediction of their physical properties possible. In 
this paper, we focus on the effects of screening of intermicellar interactions, achieved 
by adding suitable amounts of the salt NaCl to CTAT samples, on the linear and nonlinear rheology of the viscoelastic gel phase. The most interesting feature of this work is 
the existence of an anomalous regime of salt (NaCl) concentration where the high 
frequency plateau modulus $G_{\circ}$ {\it increases} on the addition of salt. The maximum 
value of $G_{\circ}$ occurs when $\frac{[NaCl]}{[CTAT]} \sim$ 1.42, where [NaCl] and 
[CTAT] represent the molarities of NaCl (42 mM) and CTAT (60mM), respectively. 
Interestingly, even as the slope of $G_{\circ}$ changes sign on increasing salt 
concentration, there is no change in the slope of the relaxation time $\tau_{R}$, while 
the slope of the zero-shear viscosity $\eta_{\circ}$ plotted {\it vs.} salt concentration 
shows a significant decrease.

\noindent Theoretical and experimental studies on the effect of electrostatics on the growth 
of cylindrical micelles in the semi-dilute regime show a much stronger concentration 
dependence of the dynamical properties in salt-free conditions than in the highly screened 
case. Kern {\it et al.} have studied the effect of an increase in the concentration of 
the dimeric gemini surfactant on the micellar size, in the absence of added salt \cite{kern}. 
On increasing surfactant concentration $c$ beyond the overlap concentration $c^{\star}$, 
the zero shear viscosity $\eta_{\circ}$ shows a very strong increase. In addition to this, 
the high frequency plateau modulus $G_{\circ}$ shows a stronger concentration dependence than 
in the highly screened case ($G_{\circ} \sim c^{3}$ for unscreened micelles, in contrast to 
the prediction $G_{\circ} \sim c^{\frac{9}{4}}$ for highly screened, flexible 
micelles) \cite{kern}. Experiments on cetyl trimethylammonium chloride (CTAC) in the presence 
of sodium salicylate (NaSal) show a tendency towards a mono-exponential stress 
relaxation process with increase in the surfactant concentration $c$, or a decrease in 
the temperature $T$ \cite{kern1}. In these experiments, the scaling relations for the 
zero-shear viscosity $\eta_{\circ}$ and the high frequency plateau modulus $G_{\circ}$ are 
given by $\eta_{\circ} \sim c^{1.1}$ and $G_{\circ} \sim c^{1.7}$, where $c$ is 
surfactant (CTAC) concentration. These scaling relations are found to be different from 
the theoretical model of Cates \cite{cates}, which predicts $\eta_{\circ} \sim c^{3.7}$ 
and $G_{\circ} \sim c^{2-2.3}$. These anomalies are explained in terms of the 
non-uniform distribution of the bound chloride and salicylate ions in the end-caps and along 
the lengths of the cylindrical micelles \cite{kern1}. These experimental results 
prompted MacKintosh {\it et al.} \cite{mackintosh} to analytically study the effect 
of electrostatics on the growth of cylindrical micelles.  Considering the interaction 
of counterions with each other, the self-energy of the free charges and the interaction of 
the counterions with the surface charge density to be the main contributions to 
the electrostatic repulsion between micelles, Mackintosh {\it et al.} find that 
charged, unscreened micelles exhibit three regimes of micellar growth with a tendency for 
very rapid growth in the semi-dilute regime. Screened micelles, on the other hand, show a 
simple and more gradual power-law growth in the semi-dilute regimes. 

\subsection{Theoretical models and some experimental results}

In contrast to polymer solutions, micellar systems have a molecular weight distribution (MWD) 
at thermal equilibrium \cite{ch5cates}. In generalizing the theory of polymers to describe 
the equilibrium properties of self-assembling cylindrical micelles (also called {\it 
living polymers} or {\it wormlike micelles}), care has to be taken to ensure a realistic 
length or molecular weight distribution, which may be done within the framework of the 
Flory-Huggins theory \cite{flory}.

\subsubsection{The reptation-reaction theory}

The dynamics of polymeric macromolecules is well understood in terms of the reptation 
model \cite{flory,doi}. Reptation dynamics implies the curvilinear diffusion of a polymer 
chain through an imaginary tube along its own contour. In this model, it is assumed that 
the stress relaxation in polymers on the application of a strain results in the deformation 
of the chain, so that the polymer reptates out of the chain to occupy a new tube which is 
in equilibrium and carries no strain \cite{doi}. For a monodisperse system, the fraction of 
the stress $G(t)$ remaining at time $t$ due to the application of a strain at $t$ = 0 is 
given by \cite{ch5cates}

\begin{equation}
G(t) = \frac{8}{\pi^{2}} \sum_{p=odd} p^{-2} \exp(\frac{-t p^{2}}{\tau_{rep}})
\end{equation}
which implies a decay of the stress with a dominant relaxation time $\tau_{rep} 
\approx \frac{{\bar L}^{2}}{D_{c}}$, $D_{c}$ being the diffusion coefficient of the 
reptating chain and $\bar L$ its average length.

\noindent In the case of living polymers with an equilibrium MWD, the contribution due to 
the breaking and recombination of chains on a time scale $\tau_{break}$ should also be 
accounted for. If $\tau_{break} >> \tau_{rep}$, the relaxation function may be written 
as \cite{cates}

\begin{equation}
G (t) \sim \exp[-constant * ({\frac{t}{\tau_{rep}}})^{\frac{1}{4}}]
\end{equation}
which indicates a non-exponential stress decay. 

\noindent Experimentally, a purely exponential stress relaxation has been found in 
many surfactant solutions \cite{rehage,ch5cates,kern,kern1} where the dynamics 
obey $\tau_{break} << \tau_{rep}$. In such cases, the relaxation spectrum shows a 
single-exponential behavior and the viscoelastic moduli of the material are given by the 
well-known Maxwell model \cite{ch5cates,fischer}
\begin{eqnarray}
G^{\prime}(\omega) = G_{\circ} \frac{\omega^{2} \tau_{R}^{2}}{1 + \omega^{2} \tau_{R}^{2}}\\
G^{\prime\prime}(\omega) = G_{\circ} \frac{\omega \tau_{R}}{1 + \omega^{2} \tau_{R}^{2}}
\end{eqnarray}
where $G^{\prime}(\omega)$ and $G^{\prime\prime}(\omega)$ are the elastic and viscous 
moduli respectively, and the complex modulus $G^{\star}(\omega)= G^{\prime}(\omega) 
+ iG^{\prime\prime}(\omega)$. In Eqns. 3 and 4, $G_{\circ}$ is the high frequency 
plateau modulus, $\omega$ is the angular frequency of the applied oscillatory stress  
and $\tau_{R}$ is a characteristic relaxation time of the sample given 
by 
$\tau_{R} = (\tau_{rep}\tau_{break})^{\frac{1}{2}}$. The complex 
viscosity $\eta^{\star}(\omega)$ may be written in terms of $G^{\prime}(\omega)$ 
and $G^{\prime\prime}(\omega)$ as follows 
\begin{equation}
\eta^{\star}(\omega) = \frac{\sqrt{G^{\prime}(\omega)^{2} + G^{\prime\prime}(\omega)^{2}}}{\omega}
\end{equation}
For viscoelastic gels, the complex viscosity is given by \cite{fischer}
\begin{equation}
|\eta ^{\star}(\omega)| = \frac{\eta_{\circ}}{\sqrt{1 + \omega^{2} \tau_{R}^{2}}} 
\end{equation}
In the Maxwell model, the zero-shear viscosity $\eta_{\circ}$ is given by 
\begin{equation}
\eta_{\circ} = G_{\circ} \tau_{R}
\end{equation}
For a distribution of relaxation times, the stress relaxation may be fitted to a 
stretched exponential model given by 
\begin{equation}
G(t) \sim \exp[-constant * ({\frac{t}{\tau_{rep}}})^{\alpha}]
\end{equation}
where the exponent $\alpha$ depends on the surfactant concentration, salinity and 
temperature, and tends to 1 when $\tau_{break} << \tau_{rep}$. In the frequency domain, 
the corresponding form is the empirical Cole Davidson model \cite{cole}, where the 
complex relaxation $G^{\star}(\omega)$ can be described as 
\begin{equation}
G^{\star}(\omega) = G_{\circ}[1 - \frac{1}{{(1 + i \omega \tau_{R})}^{\alpha}}]
\end{equation}
where $G_{\circ}$ corresponds to $G(\omega \rightarrow \infty)$. $G_{\circ}$ can be used 
to estimate $\xi$, the correlation length of concentration fluctuations, by using the 
relation $G_{\circ} \sim \frac{k_{B}T}{\xi^{3}}$ \cite{ch5cates,candau1}.

\subsubsection{The Giesekus model}

The Giesekus model \cite{giesekus} can be used to predict the nonlinear flow properties 
of surfactant solutions. This model uses the reptation theory as the starting point 
and introduces a deformation-dependent mobility tensor $\beta$ in order to account for 
the orientation effects of flow. Using a linear dependence between $\beta$ and the configuration tensor $C$, and combining this equation with the upper-convected Maxwell equation 
$\beta \sigma = \tau_{R} \frac{\partial \sigma}{\partial t} = 2 \eta \dot\gamma $, 
where $\sigma$ is the shear stress, $\dot\gamma$ is the shear rate and $\tau_{R}$ is 
the characteristic relaxation time, Giesekus obtained the following relations 
between rheological parameters such as the shear stress $\sigma$, and the first and 
second normal stress differences $N_{1}$ and $N_{2}$ in the $t \rightarrow \infty$ limit: 
\begin{eqnarray}  
\sigma (\infty , \dot\gamma) = \frac{G_{\circ}}{2 \tau_{R} \dot\gamma}(\sqrt{1 + 4 \tau_{R}^{2} \dot\gamma^{2}}-1)\\
N_{1} (\infty , \dot\gamma) = 2 G_{\circ} {\frac{1 - \Lambda^{2}}{\Lambda}}\\
N_{2} (\infty , \dot\gamma) =  G_{\circ} (\Lambda -1)
\end{eqnarray}
where $\Lambda ^{2} = 
\frac{\sqrt{1 + 4 \tau_{R}^{2} \dot\gamma^{2}}-1}{2 \tau_{R}^{2} \dot\gamma^{2}}$. In 
the Giesekus model, the shear viscosity $\eta(\dot\gamma)$ is given by 
\begin{equation}
\eta (\dot\gamma) = \frac{\eta_{\circ}}{2 \tau_{R}^{2} \dot\gamma^{2}}(\sqrt{1 + 4 \tau_{R}^{2} \dot\gamma^{2}}-1)
\end{equation}
\subsubsection{Cox-Merz rule}

The Cox-Merz rule is a semi-empirical rule which relates the complex 
viscosity $\eta^{\star}(\omega)$ with the shear viscosity $\eta (\dot\gamma)$ in the 
following way : 
\begin{equation}
\eta(\dot\gamma ) = |\eta ^{\star} (\omega)|, \hspace{1cm} \omega = \dot\gamma
\end{equation}
The shear viscosity $\eta(\dot\gamma)$ is predicted by the Giesekus model and is given by 
Eqn. 13. The dynamic viscosity may be calculated from the viscoelastic moduli using Eqn. 5 
and may be fitted to Eqn. 6. For the CTAB/ NaSal (60mM/ 350mM) system forming giant 
wormlike micelles, a relatively good agreement to the Cox-Merz rule is observed \cite{fischer}. 

\subsubsection{Connected micelles}
For suitably high concentrations of the added electrolyte, many of the measured 
rheological properties of entangled micellar solutions are found to deviate considerably 
from the predictions of the reptation-reaction theory \cite{ch5cates}. Unusually high 
fluidity has been observed in aqueous micellar solutions of the system cetylpyridinium 
chlorate (CPClO$_{3}$) / sodium chlorate (NaClO$_{3}$) \cite{appell}. For the 
system hexadecyltrimethylammonium bromide (CTAB)/ potassium bromide (KBr)/ water, Khatory 
{\it et al.} have observed anomalous scaling of the zero shear viscosity $\eta_{\circ}$ and 
the high frequency plateau modulus $G_{\circ}$ with surfactant concentration \cite{khatory}. 
The zero-shear viscosity $\eta_{\circ}$ of cetyltrimethylammonium chloride (CTAC) and 
sodium salicylate (NaSal) shows a peak at $\frac{[NaSal]}{[CTAC]}\sim 0.6$ \cite{ali}. 
The increase and decrease of $\eta_{\circ}$ obey power-law scaling relations with 
CTAC concentration, characterized by exponents 1.1 and -2.1 respectively. A peak in the 
zero-shear viscosity has also been observed in aqueous solutions of CPyCl/NaSal 
at $\frac{[NaSal]}{[CPyCl]} \sim$ 1 \cite{rehage1}. These results cannot be explained in 
terms of the theory for entangled micelles \cite{cates}, but can be understood by 
considering intermicellar branching at high salt concentrations. These connections 
between micelles characterize a new relaxation process which involves the sliding of 
connections along the micelles \cite{drye,lequeux}. The general features of stress 
relaxation seen in linear micelles are preserved in the case of branched micelles if 
one replaces the average length $\bar L$ by a new length $\bar L_{c}$, where 
$\bar L_{c} = {\frac{n_{2}}{n_{1}+2n_{3}}}l_{p}$ \cite{khatory,lequeux}.  In the expression 
for $\bar L_{c}$, $l_{p}$ is the persistence length, $n_{1}$ is the concentration of the 
end caps, $n_{2}$ is the number density of the persistence lengths and $n_{3}$ is the 
number density of 3-fold network junctions. This model gives rise to a relaxation process 
that is faster than the predictions of the theory for reptation and reversible 
scission \cite{cates}, and explains the anomalously high fluidity seen in some systems 
of wormlike micelles at high salt or surfactant concentrations \cite{khatory,ali}. 
%\newpage
\section{Sample preparation and apparatus used}

Samples of CTAT/ NaCl/ water were prepared by weighing out the requisite quantity of CTAT in 
a microbalance and  dissolving it in brine (solution of NaCl in deionized and distilled 
water) prepared at the following concentrations: 0 mM, 0.5mM, 1mM, 2mM, 5mM, 10mM, 20mM, 
40mM, 60mM, 80mM, 90mM, 100mM and 120mM. The concentration of CTAT in the brine solution 
was kept constant at 42mM (1.9 wt.\%). The samples were kept in an incubator at 
30$^{\circ}$C for a week and shaken frequently to ensure homogenization. All the 
experiments reported below were conducted at a fixed temperature of 25$^{\circ}$C. 
The oscillatory and flow measurements were performed in an AR 1000N stress controlled 
rheometer (T. A. Instruments, U. K.), using a cone-and-plate geometry of radius 4 cm and 
angle 1$^{\circ}$59'. Sponges were used as solvent traps to prevent evaporation of the 
solvent (water) during the experiment.

\section{Experimental results}
In this section, we first describe the linear rheology results (section 3.1), followed by 
the nonlinear rheology results (section 3.2) and attempt to correlate the results obtained 
from the two, with an aim to understanding the flow behavior of CTAT in the presence of 
salt (section 4).
\subsection{Linear Rheology}
\noindent Fig. 1 shows the frequency response ($G^{\prime}(\omega)$ 
and $G^{\prime\prime}(\omega)$ {\it vs.} $\omega$) measurements for CTAT/ NaCl/ water for a 
few typical salt concentrations ($c_{NaCl}$ = (a) 0mM, (b) 20mM, (c) 60mM and (d) 100mM) 
over three decades of angular frequencies, and the corresponding fits to the real and 
imaginary parts of the Cole-Davidson form given by Eqn. 9 (shown by solid lines in each plot). 

\noindent The fits to the Cole-Davidson form of the frequency response data in the presence 
of 20 - 120mM NaCl yield $\alpha \sim$ 0.80 - 0.90, in comparison to the unscreened 
micellar phase (0 mM NaCl) where $\alpha \sim$ 0.15. For the NaCl concentrations of 0.5mM, 
1mM, 2mM, 5mM and 10mM, the values of $\alpha$ are found to increase with increasing 
molarity 
of the brine solution. Fig. 2 shows the values of $\alpha$ obtained from the Cole-Davidson 
fit as a function of salt concentration, which indicates a gradual crossover from 
non-exponential to single-exponential stress relaxation on the addition of NaCl to 
CTAT solutions. Similar observations by Rehage {\it et al.} \cite{rehage,rehage1} on 
adding NaSal to CPyCl solutions have been explained in terms of a crossover from 
diffusion-controlled to kinetically-controlled stress relaxation processes. 
The fits shown in Fig. 1 also give us estimates of the high frequency plateau 
modulus $G_{\circ}$ and the terminal relaxation time $\tau_{R}$.  The values of $\tau_{R}$ 
have also been calculated from the crossover frequency $\omega_{co}$ using the 
relation $\tau_{R} = \frac{1}{\omega_{co}}$ (Fig. 7(a)). The data acquired in the presence of 
20 - 120mM NaCl can also be fitted satisfactorily to the Maxwell model \cite{ch5cates} as 
is seen from the semicircular nature of the Cole-Cole plots ($\frac{G^{\prime}(\omega)}{G_{\circ}}$ versus  
$\frac{G^{\prime\prime}(\omega)}{G_{\circ}}$ plotted in Fig. 3). $G_{\circ}$ is the value of 
the high frequency plateau modulus obtained from fits to the Cole-Davidson model. 

\noindent The Cole-Cole plots for the CTAT samples with low salt content ($\le$ 5mM) are 
found to deviate considerably from the semicircular behavior characteristic of a 
Maxwellian fluid. We note that these deviations at high frequencies 
($\omega \ge \omega_{b}$) and the subsequent upturn ($\omega \ge \omega_{e}$) in the 
Cole-Cole plots are due to the reversible breaking of the wormlike chains at a 
characteristic time $\tau_{break}$ and  the contribution of the localized dynamics 
between entanglements (the Rouse modes), respectively \cite{granek}. The absence of a 
well-defined plateau in the elastic modulus $G^{\prime}(\omega)$ of the samples with low 
salt content at high angular frequencies $\omega$, is possibly due to long breaking times or 
the occurrence of Rouse modes at relatively low frequencies. 

\noindent Fig. 4 shows the plots of the complex viscosities $\eta^{\star}(\omega)$ 
{\it vs.} $\omega$ measured for the samples with $c_{NaCl}$ equal to (a) 0mM, (b) 20mM, (c) 
60mM and (d) 100mM, {\it vs.} the angular frequencies $\omega$, and the corresponding 
fits 
to the model for giant wormlike micelles given in Eqn. 6 \cite{fischer}. The fits, shown 
by solid lines, are found to be poor for the unscreened case, but agree very well with 
the experimental data for the samples with high concentrations of added salt ($c_{NaCl} 
\ge$ 20mM). The values of $\eta_{\circ}$ obtained from these fits are shown in Fig. 7(c) 
by solid triangles.
\begin{table}
\begin{center}
%\label{Table 5.1}
\caption{The parameters $\tau_{R}$ and $\eta_{\circ}$, obtained from the fits 
of $\eta^{\star}(\omega)$ {\it vs.} $\omega$ to Eqn. 6 (denoted by $DV$) and from the fits 
of $\eta(\dot\gamma)$ {\it vs.} $\dot\gamma$ to the Giesekus model (Eqn. 13, denoted by 
$GM$) are tabulated below :}
\vspace{0.5cm}
\begin{tabular}{|c|c|c||c|c|}\hline 
{\rule[-3mm]{0mm}{8mm} NaCl (mM)} &$\tau_{R}$ s(DV) &$\tau_{R}$ s (GM) &$\eta_{\circ}$ Pa-s (DV) &$\eta_{\circ}$ Pa-s (GM)\\  \hline \hline
{\rule[-3mm]{0mm}{8mm} 0}& 8.78  &12.88 &42.42 &36.69\\ \hline
{\rule[-3mm]{0mm}{8mm} 0.5}&13.93  &21.03 &49.62 &36.72\\ \hline
{\rule[-3mm]{0mm}{8mm} 1}&15.16  &24.5 &62.13 &51.02\\ \hline
{\rule[-3mm]{0mm}{8mm} 2}&15.77  &22.35 &80.27 &70.43\\ \hline
{\rule[-3mm]{0mm}{8mm} 5}&36.69 &31.61 &42.42 &39.05\\ \hline
{\rule[-3mm]{0mm}{8mm} 10}&6.00 &7.2 &45.52 &46.08\\ \hline
{\rule[-3mm]{0mm}{8mm} 20}& 3.53  &3.44 &38.7 &44.35\\ \hline
{\rule[-3mm]{0mm}{8mm} 40}& 2.30  &1.77 &32.49 &22.21\\ \hline
{\rule[-3mm]{0mm}{8mm} 60}&1.79 &1.46 &27.98 &29.63\\ \hline
{\rule[-3mm]{0mm}{8mm} 80}&1.51  &1.06 &17.09 &15.24\\ \hline
{\rule[-3mm]{0mm}{8mm} 90}&1.30  &0.98 &11.14 &11.72\\ \hline
{\rule[-3mm]{0mm}{8mm} 100}&1.21  &0.58 &9.50 &6.74\\ \hline
{\rule[-3mm]{0mm}{8mm} 120}&1.08  &0.50 &6.56 &6.03\\ \hline
%$\tau_{R} (CD)$ \\ \hline
%$\tau_{R} (GG')$$ \\ \hline
\end{tabular}
\end{center}
\end{table} 
\subsection{Nonlinear rheology} 
Nonlinear rheology measurements have been used to estimate the zero-shear 
viscosity $\eta_{\circ}$ and the relaxation time $\tau_{R}$ from the flow curves. Here the 
shear stress $\sigma$ and the viscosity $\eta$ are measured simultaneously as a function of 
the shear rate $\dot\gamma$. Fig. 5 shows the fits (shown by solid lines) to the Giesekus 
model (Eqn. 13) of the shear viscosities $\eta(\dot\gamma)$ (indicated by circles) 
{\it vs.} shear rates $\dot\gamma$ for the samples with salt content (a) 0mM, (b) 20mM, (c) 
60mM and (d) 100mM respectively. For high salt concentrations, the fits to the Giesekus 
model are found to deviate considerably from the  experimental data at high values of $\dot\gamma$.
The values of $\eta_{\circ}$ and $\tau_{R}$ obtained from these fits are compared to 
those obtained from the fits to $\eta^{\star}(\omega)$ (Table 1). The values thus 
calculated from the linear and nonlinear rheology measurements deviate considerably from 
each other at low salt concentrations but show some agreement at $c_{NaCl} \ge$ 20mM. 

\section{Discussions}

In this section, we will attempt to quantify and understand the changes in the 
viscoelastic parameters of the CTAT/ NaCl/ water samples due to the increased screening 
of intermicellar interactions as a result of the addition of salt. 

\subsection{Applicability of the Cox-Merz rule}

\noindent As discussed in section 1.1.3, viscoelastic gels usually follow the empirical 
Cox-Merz rule \cite{fischer}. To see this in the present case, we have plotted in Fig. 6 
the normalized viscosities $\frac{\eta^{\star}(\omega)}{\eta_{\circ}}$ 
and $\frac{\eta(\dot\gamma)}{\eta_{\circ}}$ versus angular frequency $\omega$ and shear 
rate $\dot\gamma$ respectively for $c_{NaCl}$ equal to (a) 0mM, (b) 20mM, (c) 60mM and 
(d) 100mM. The curves show an excellent superposition at $c_{NaCl}$=
60mM ($\frac{[NaCl]}{[CTAT]}$=1.42) over the entire range of $\omega$ and $\dot\gamma$. 
We notice significant deviations in the values of 
$\frac{\eta^{\star}(\omega)}{\eta_{\circ}}$ and $\frac{\eta(\dot\gamma)}{\eta_{\circ}}$ at 
high $\omega$ and $\dot\gamma$ at very low and very high salt concentrations. 

\subsection{Existence of three regimes of contrasting flow behaviors}

\noindent Fig. 7(a) shows the relaxation time $\tau_{R}$, obtained using the relation 
$\tau_{R} \sim \omega_{co}^{-1}$, {\it vs.} $c_{NaCl}$. On increasing $c_{NaCl}$, 
$\tau_{R}$ shows an initial increase, followed by a strong decrease. The dependence 
of $\tau_{R}$ on $c_{NaCl}$ in these regimes may be fitted to the relation $\tau_{R} 
\sim c_{NaCl}^{\beta}$ (shown by solid lines in Fig. 7(a)). The values of $\beta$ is +0.11 
for $c_{NaCl} \le$ 2mM (regime I, $\frac{[NaCl]}{[CTAT]} \le$ 0.05), and changes to 
-0.65 thereafter. 

\noindent Fig. 7(b) shows the plots of the high frequency plateau modulus $G_{\circ}$ 
obtained in the following ways: the values obtained from the Cole-Davidson fits (solid 
circles), and  the values of $G^{\prime}(\omega)$ at a high frequency $\omega$ = 
29 rads$^{-1}$ (solid triangles). The two sets of values of $G_{\circ}$ deviate considerably 
at $c_{NaCl} \le$ 2mM (regime I), but agree very well at higher salt concentrations 
($c_{NaCl} \ge$5mM).  At 5mM $\le c_{NaCl} \le$ 60mM, $G_{\circ}$ increases with 
salt concentration, followed by a decrease at $c_{NaCl} \ge$ 60mM. These two 
regimes, characterized by different slopes (shown in Fig. 7(b) and marked as regimes II 
and III), have been fitted to $G_{\circ} \sim c_{NaCl}^{\beta^{\prime}}$, 
where $\beta^{\prime}$=0.31 in regime II and -1.38 in regime III. 

\noindent The same three regimes can be very clearly distinguished in Fig. 7(c), 
where $\eta_{\circ}$ is plotted {\it vs.} $c_{NaCl}$. The values of $\eta_{\circ}$ are 
obtained from the fits of $\eta^{\star}$ (solid triangles) and $\eta(\dot\gamma)$ 
(solid circles) discussed above. $\eta_{\circ}$ shows an initial increase in regime I, 
followed by a weak decrease in regime II and subsequently, a much stronger decrease in 
regime III. In all the three regimes, $\eta_{\circ}$ has been fitted to the power 
law $\eta_{\circ} \sim c_{NaCl}^{\beta^{\prime\prime}}$, where $\beta^{\prime\prime}$ is 
equal to 0.35 in regime I (fit shown by dashed line), -0.29 in regime II (fit shown by 
solid line) and -2.04 in regime III (fit shown by dashed line). The weak increase 
in $\eta_{\circ}$ at low $c_{NaCl}$ may be explained in terms of micellar growth as a result 
of enhanced screening of intermicellar interaction on the addition of NaCl. The 
subsequent decrease in $\eta_{\circ}$ at high $c_{NaCl}$ may be explained in 
terms 
of intermicellar connections \cite{rehage,khatory,ali}. We would like to note here 
that fluorescence recovery after fringe pattern photobleaching (FRAPP) experiments have 
also shown an increase in the diffusion coefficients ({\it i.e.} a decrease in the 
effective viscosity) of aqueous, semi-dilute  solutions of CTAT micelles in the presence of 
0.1 M and 1 M NaCl with increasing CTAT concentrations \cite{narayanan}. This effect 
increases with the increase in the salt concentration and has been explained in terms 
of connected micelles. 
These intermicellar connections serve as sliding contacts, aiding the faster reptation of 
the micelles, and hence decreasing the viscosity of the system \cite{lequeux}. An 
interesting feature of our data is the existence of two distinct regimes (regimes II and III 
in Fig. 7(c)) where $\eta_{\circ}$ shows a decrease with $c_{NaCl}$. In contrast to 
previous experiments with gemini surfactants \cite{kern}, there is a regime of weak decrease 
of $\eta_{\circ}$ (regime II) followed by a much stronger decrease on increasing 
$c_{NaCl}$ (regime III). At the cross-over between regimes II and III ($c_{NaCl}$ = 60mM), 
the change in the slope of $\eta_{\circ}$ (Fig. 7(c)) is accompanied by a change in the sign 
of the slope of $G_{\circ}$, as shown in Fig. 7(b). In regime II, $G_{\circ}$ shows an 
anomalous increase with $c_{NaCl}$. 

\noindent The addition of salt is known to encourage micellar growth \cite{ch5cates}. 
However, as the CTAT concentration is fixed and lies in the semi-dilute regime, we do not 
expect $\xi$, the correlation length of concentration fluctuations to change appreciably. 
For flexible micelles, the correlation length $\xi$ is related to $G_{\circ}$ as $G_{\circ} 
\sim \frac{k_{B}T}{\xi^{3}}$ \cite{ch5cates}. The plateau modulus $G_{\circ}$ may be written 
as $G_{\circ} \sim \frac{ck_{B}T}{l_{e}}$, where $c$ is the macromolecule concentration 
and $l_{e}$ is the entanglement length \cite{flory}. $l_{e}$ scales with the persistence 
length $l_{p}$ and the correlation length $\xi$ as 
$\l_{e} \sim \frac{\xi^\frac{5}{3}}{l_{p}^\frac{2}{3}}$. An increase in $G_{\circ}$, 
therefore, implies a decrease in entanglement length $l_{e}$ and an increase in the 
persistence length $l_{p}$. An increase in $l_{p}$ is indicative of an increase in the 
surface charge of the micelles. Such an increase in the surface charge could occur due to 
the unbinding of the tosylate counterions from CTA$^{+}$ on the addition of NaCl 
\cite{private}. Low-frequency conductivity measurements have confirmed the phenomenon 
of counterion unbinding in the liquid crystalline phase of cesium perfluoro-octanol 
(CsPFO)/ PFO/ water on the addition of alcohol \cite{li}. However, at $c_{NaCl} \ge$ 
60mM (regime III), the micelles are again screened completely, as indicated by the decrease 
in $G_{\circ}$ and $\eta_{\circ}$ in this regime. We would like to note here that the 
decrease in viscosity of the system CTAC/NaSal with increasing concentrations of CTAC 
follows the relation $\eta_{\circ} \sim c_{NaCl}^{\beta^{\prime\prime}}$, 
where $\beta^{\prime\prime}$ = -2.1 \cite{ali}, very close to the value  
$\beta^{\prime\prime}$ = -2.04 obtained by us in regime III.

\subsection{Superposition of the linear and nonlinear rheology data for CTAT/ NaCl/ water}

\noindent In this section, we discuss the superposition of the frequency response and 
flow curves of the CTAT/ NaCl/ water samples. In Figs. 8 and 9, we have plotted master curves 
by normalizing the viscoelastic parameters by suitable quantities. The elastic and 
viscous moduli are normalized by $\frac{\eta_{\circ}}{\tau_{R}}$, the dynamic 
viscosity $\eta^{\star}$ by $G_{\circ}\tau_{R}$, the shear stress $\sigma$ by $G_{\circ}$ 
and the shear rate $\dot\gamma$ by $\tau_{R}^{-1}$, respectively. Similar master curves for 
the linear rheology data of CTAT have been observed by Soltero {\it et al.} \cite{soltero2} 
on increasing the concentration of CTAT. In Fig. 8, (a) and (b) correspond to the plots of 
the normalized $G^{\prime\prime}(\omega)$ and $G^{\prime}(\omega)$ data respectively, while 
(c) shows the plots of the normalized dynamic viscosity $\eta^{\star}(\omega)$. The curves 
for the normalized $G^{\prime\prime}(\omega)$ at 0mM NaCl are found to deviate considerably 
from the master curve on which the data corresponding to $c_{NaCl} \ge$ 20 mM lie, which 
is indicative of a large difference in the viscous flow properties of CTAT in the screened 
and unscreened limits. 

\noindent It is also possible to superpose the normalized stress $\frac{\sigma}{G_{\circ}}$ 
{\it vs.} the  normalized shear rate $\dot\gamma\tau_{R}$ plots of the CTAT/ NaCl/ 
water samples. All the flow curves are found to superpose very well in the regime of 
low $\dot\gamma\tau_{R}$ (Newtonian regime, characterized by unit slope of $\sigma$ 
{\it vs.} $\dot\gamma$ on a log-log plot), but show  deviations at higher values 
of $\dot\gamma\tau_{R}$ (Fig. 9). The normalized shear stress is found to level off to a 
plateau at $\dot\gamma\tau_{R} \sim$ 1, with the slope of the plateau increasing 
monotonically with increase in salt concentrations. Berret {\it et al.} \cite{ch5berret} 
have observed perfect superposition for CPyCl /NaSal/ water samples in the Newtonian regime 
of the flow curve, while the slope of the normalized stress in the plateau (nonlinear) 
regime increases on increasing the sample temperature.  Interestingly, the branching of 
the micelles that we observe at $c_{NaCl} \ge$ 20mM does not have any significant effect on 
the master phase diagrams obtained from the linear rheology experiments (Fig. 8). The 
normalized flow curves of CTAT/ NaCl/ water, however, show a significant change in the slope 
of the 'plateau' region on increasing $c_{NaCl}$, which points to a change in the nonlinear 
flow behavior of the aqueous CTAT/ NaCl samples on the addition of salt. 

\section{Conclusions}
In this paper, we have discussed the modifications of the viscoelastic properties of 
CTAT (concentration of CTAT is kept fixed at 42mM) as a result of the addition of the salt. 
We find that the data may be divided into three distinct regimes, where $\eta_{\circ}$ 
and $G_{\circ}$ can be fitted to different power-laws. The decrease of $\eta_{\circ}$ at 
high salt concentration has been explained in terms of intermicellar branching 
\cite{lequeux}. The anomalous increase of $G_{\circ}$ on increasing salt concentration has 
been explained in terms of tosylate unbinding from the CTA$^{+}$ on the addition of salt. 
It will be interesting to understand the precise mechanism of this counterion 
unbinding phenomenon. Significantly, at $c_{NaCl}$=60mM, the persistence length of the 
micelles is maximum, which indicates the presence of highly charged micelles at 
these concentrations. Interestingly, the scaling of $\tau_{R}$ on $c_{NaCl}$ does not 
change from regime II to regime III. As shown in Fig. 6, the superposition 
between $\eta^{\star}(\omega)$ and $\eta(\dot\gamma)$ is very good at $c_{NaCl}$=
60mM. 

\noindent It is also worth noting that the branching of the micelles does not appreciably 
alter the scaling of the linear rheology parameters (plots of normalized 
$G^{\prime}(\omega)$, $G^{\prime\prime}(\omega)$ and $\eta^{\star}(\omega)$ for $c_{NaCl} 
\ge$ 20 mM, shown in Fig. 8). In contrast, when the flow curves are superposed to lie on 
a single master curve, we find that the slope of the plateau increases on increasing 
$c_{NaCl}$, indicating a modification of the shear banding properties of the samples in 
the presence of intermicellar connections. Significant changes in the nonlinear flow behavior 
of CTAB/ NaSal/ water have been observed from small angle light scattering experiments 
on increasing salicylate counterions \cite{egmond}.

\noindent In the light of the present work, it will be worthwhile to study more exhaustively 
the linear and nonlinear rheology of cylindrical micelles as a function of the surface 
charge, with and without multiconnected junctions. Anomalous flow properties of 
surfactant systems like CPyCl/ NaSal/ water have been previously observed at about 
equimolar proportions of CPyCl and NaSal \cite{rehage,rehage1}. Till date, there have 
been several studies on the effect of the chemical formula and lipophilicity of the 
counterions on the growth of a micelle \cite{magid,oda}. It will be extremely interesting 
to undertake a detailed study of the effects of the chemical nature of the counterion on 
its unbinding from a micelle. 

\section{Acknowledgements}
The authors would like to thank Prof. M. Cates for useful discussions. They would like to 
thank Profs. S. Ramaswamy, P. R. Nott and V. Kumaran for the use of the rheometer.

\newpage
\section{Figure Captions}
\vspace{0.75cm}
{\bf Figure 1:} The frequency response curves  ($G^{\prime}(\omega)$ denoted by circles 
and $G^{\prime\prime}(\omega)$ denoted by triangles versus the angular frequency $\omega$) 
of aqueous solutions of 42mM CTAT + NaCl and their corresponding fits to the Cole-Davidson 
model (shown by solid lines). The CTAT concentration is maintained constant at 42mM while 
the NaCl concentrations corresponding to the graphs are (a) 0mM, (b) 20mM, (c) 60mM, (d) 100mM. 

\vspace{0.5cm}
{\bf Figure 2:} Plot of the values of $\alpha$ obtained from the fits to the Cole-Davidson 
model {\it vs.} the salt concentration $c_{NaCl}$. The concentration of CTAT is fixed at 42mM.

\vspace{0.5cm}
{\bf Figure 3:} Normalized Cole-cole plots obtained by plotting $\frac{G^{\prime}(\omega)}{G_{\circ}}$ {\it vs.} 
$\frac{G^{\prime\prime}(\omega)}{G_{\circ}}$ for CTAT (42mM)/ NaCl/ water. The different 
symbols correspond to the following NaCl concentrations: (a) squares for 0mM, (b) 
plus-centred squares for 0.5mM, (c) plus-centred circles for 1mM, (d) dash-centred squares 
for 2mM, (e) $\times$-centred circles for 5mM, (f) circles for 20mM, (g) up-triangles for 
40mM, (h) down-triangles for 60mM, (i) plus signs for 80mM, (j) $\times$ signs for 
90mM, (k)$\star$ signs for 100mM and (l) - for 120mM.

\vspace{0.5cm}
{\bf Figure 4:} The dynamic viscosity $\eta^{\star}(\omega)$ versus $\omega$ for CTAT 
(42mM)/ NaCl/ water samples. The solid lines show the fits to Eqn. 6. The NaCl 
concentrations corresponding to each plot is as follows: (a) 0mM, (b) 20mM, (c) 60mM, (d) 
100mM.

\vspace{0.5cm}
{\bf Figure 5:} The viscosity $\eta(\dot\gamma)$ versus $\dot\gamma$ for 42mM CTAT + 
NaCl samples.  The NaCl concentrations corresponding to each plot is as follows: (a) 0mM, 
(b) 20mM, (c) 60mM, (d) 100mM. The solid lines show the fits to the Giesekus model.

\vspace{0.5cm}
{\bf Figure 6:} The normalized shear viscosity $\frac{\eta(\dot\gamma)}{\eta_{\circ}}$ 
(circles) and normalized dynamic viscosity 
$\frac{\eta^{\star}(\omega)}{\eta_{\circ}}$ (triangles) plots versus $\dot\gamma$ and 
$\omega$ for CTAT (42mM)/ NaCl/ water samples. The NaCl concentrations corresponding to 
each plot is as follows: (a) 0mM, (b) 20mM, (c) 60mM, (d) 100mM. The solid lines show the 
fits to the Giesekus model.

\vspace{0.5cm}
{\bf Figure 7:} The values of $\tau_{R}$, obtained from the crossover frequency $\omega_{co}$ 
of $G^{\prime}$ and $G^{\prime\prime}$, have been plotted in (a) {\it vs.} $c_{NaCl}$. In (b) 
we have plotted the values of $G_{\circ}$ obtained from the  fits to the Cole-Davidson 
model (indicated by filled circles) and from the values of $G^{\prime}$ at $\omega$ = 
29 rads$^{-1}$ (filled triangles). (c) shows the plots of the zero-shear 
viscosities $\eta_{\circ}$ with increasing $c_{NaCl}$, obtained from the fits to the 
dynamic viscosity $\eta^{\star}$ (solid triangles) and the fits to the Giesekus model of 
the shear viscosity $\eta$ (solid circles).  The solid and dashed lines are the fits to $A 
\sim c_{NaCl}^{\beta}$, $A = \tau_{R}, G_{\circ}$ and $\eta_{\circ}$. The values of $\beta$ 
in each regime are noted on the graphs.

\vspace{0.5cm}
{\bf Figure 8:} (a) and (b) show the values of $G^{\prime\prime}(\omega)$ 
and $G^{\prime}(\omega)$, both scaled by $\frac{\eta_{\circ}}{\tau_{R}}$ 
{\it vs.} $\omega\tau_{R}$. Apart from the measurements with 0mM NaCl (depicted by squares), 
all other curves lie on a master curve. (c) shows the scaled dynamic 
viscosity $\frac{\eta^{\star}(\omega)}{G_{\circ}\tau_{R}}$ {\it vs.} $\omega\tau_{R}$. In 
the diagrams, circles correspond to 20mM NaCl, up-triangles for 40mM, down-triangles for 
60mM, plus signs for 80mM, cross signs for 90mM, stars for 100mM and bars for 120mM.

\vspace{0.5cm}
{\bf Figure 9:} The normalized flow curves of CTAT (42mM)/ NaCl/ water, where $\sigma$ 
and $\dot\gamma$ are scaled as  $\sigma \rightarrow \frac{\sigma}{G_{\circ}}$ and 
$\dot\gamma \rightarrow \dot\gamma\tau_{R}$. In the diagrams, squares stand for 0mM NaCl 
while the keys for the other symbols are the same as in Fig. 8.

\newpage
\begin{figure}[h]
\centerline{\epsfxsize = 14cm \epsfysize = 14cm \epsfbox{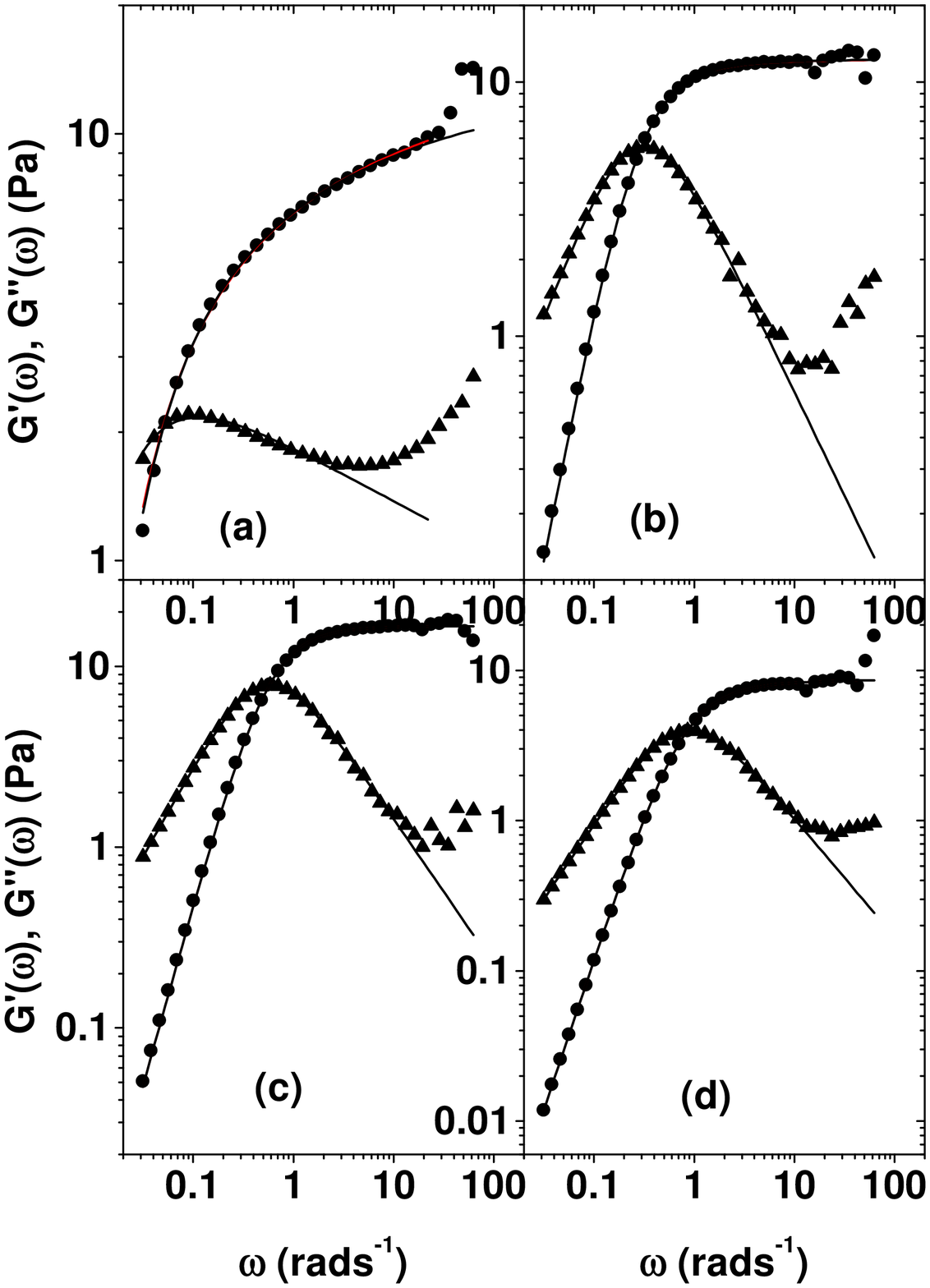}}
\end{figure}
\centerline{Figure 1}

\newpage
\begin{figure}[h]
\centerline{\epsfxsize = 16cm \epsfysize = 12cm \epsfbox{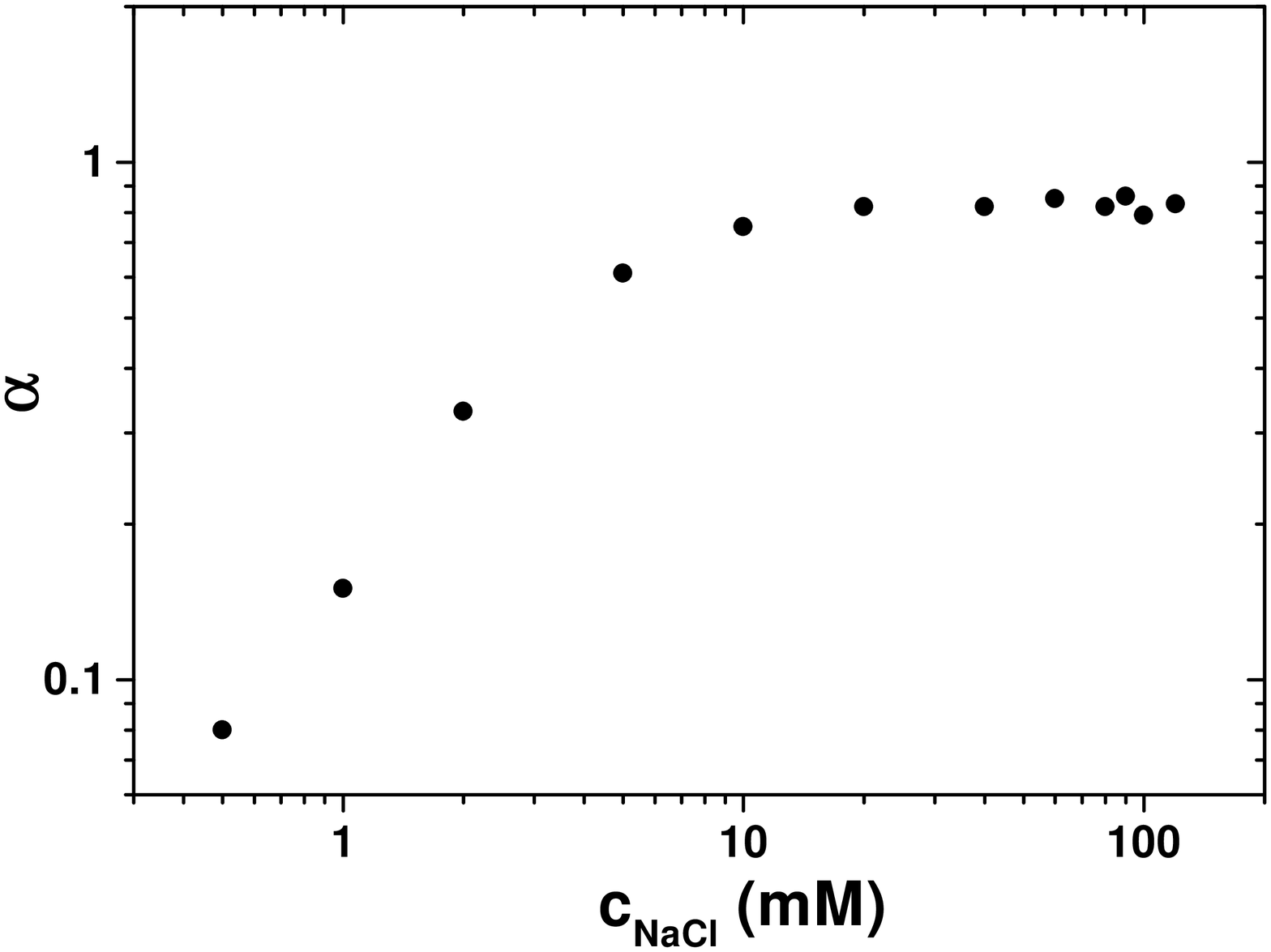}}
\end{figure}
\centerline{Figure 2}

\newpage
\begin{figure}[h]
\centerline{\epsfxsize = 14cm \epsfysize = 12cm \epsfbox{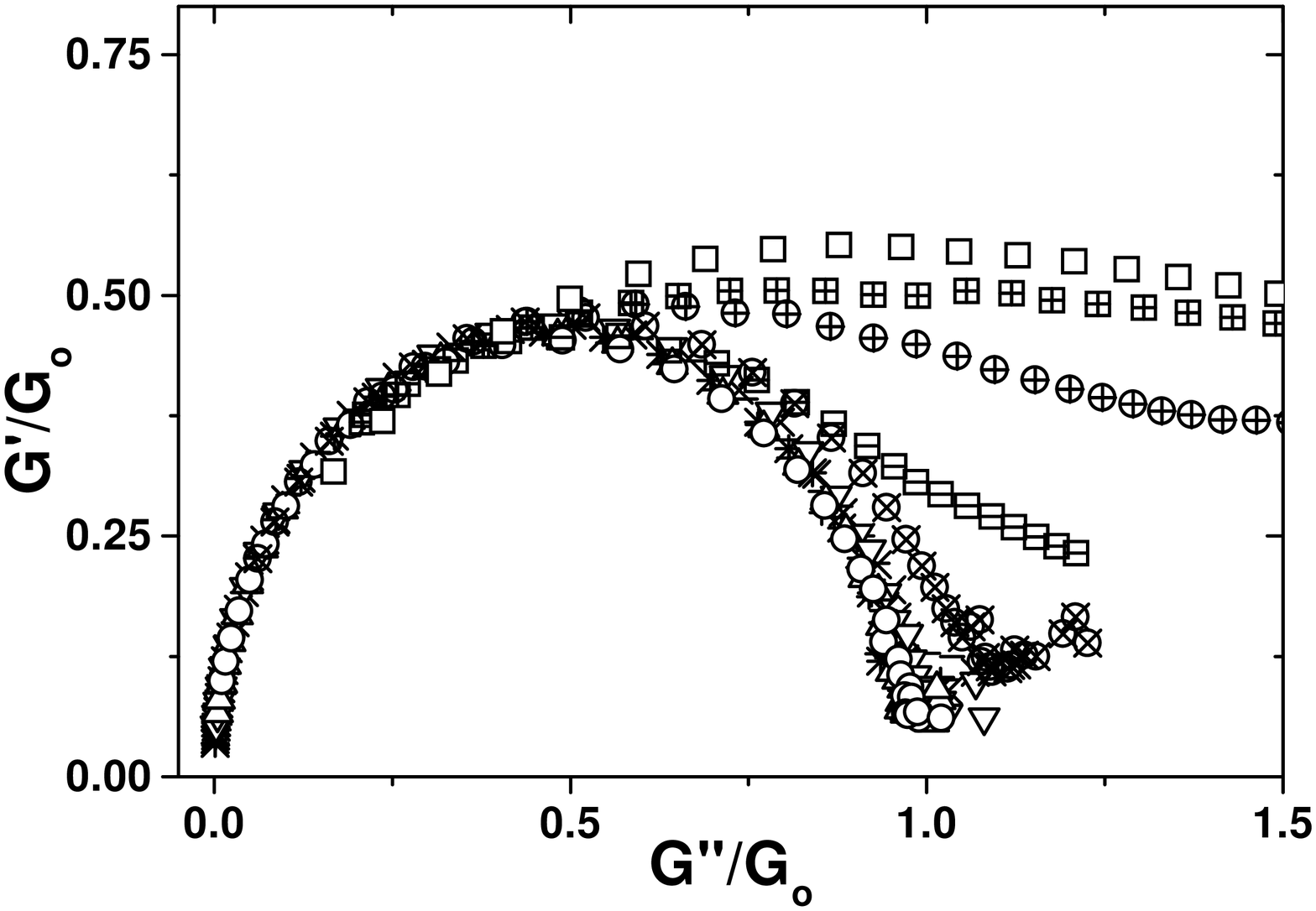}}
\end{figure}
\centerline{Figure 3}

\newpage
\begin{figure}[h]
\centerline{\epsfxsize = 12cm \epsfysize = 12cm \epsfbox{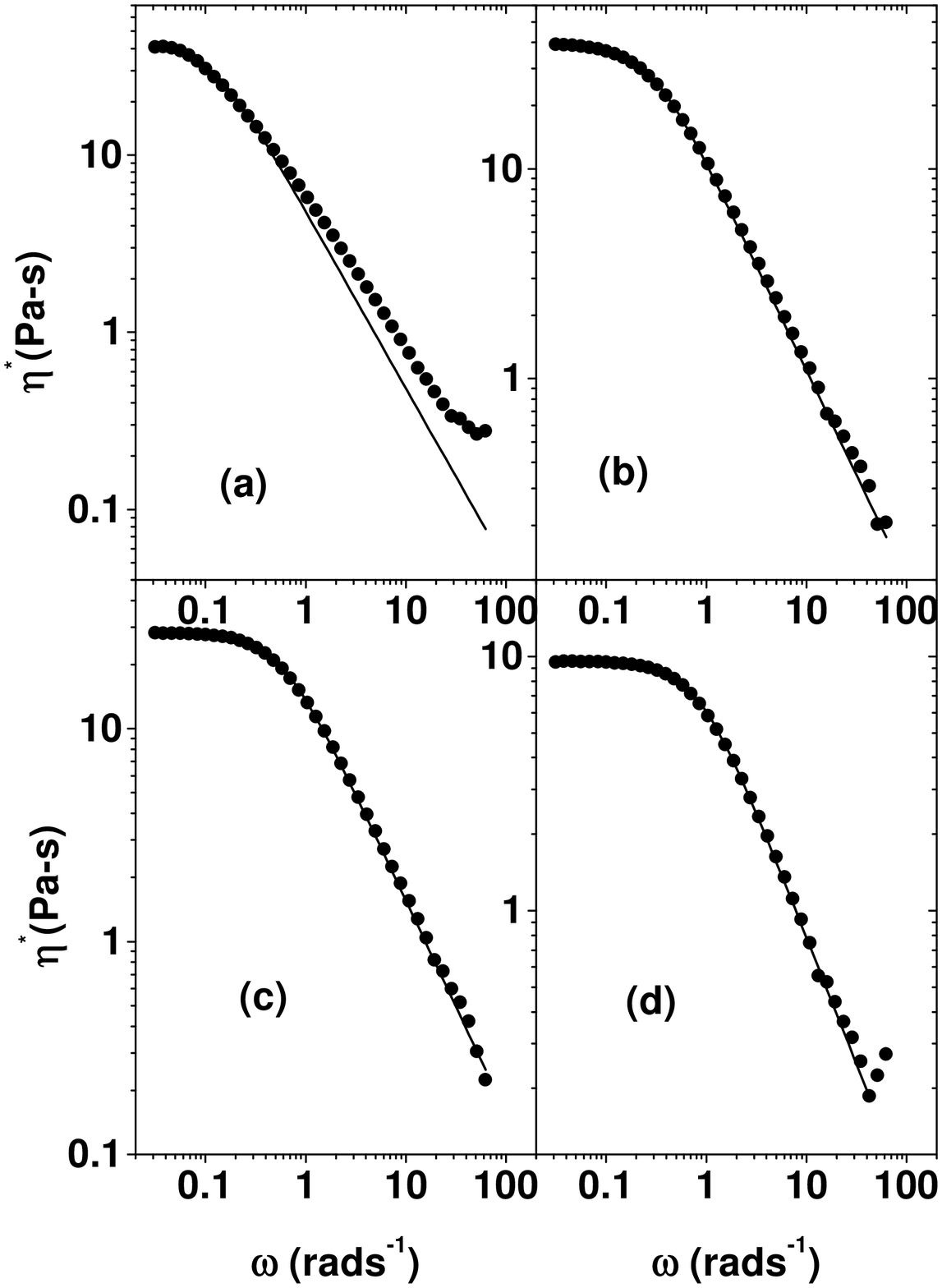}}
\end{figure}
\centerline{Figure 4}

\newpage
\begin{figure}[h]
\centerline{\epsfxsize = 12cm \epsfysize = 12cm \epsfbox{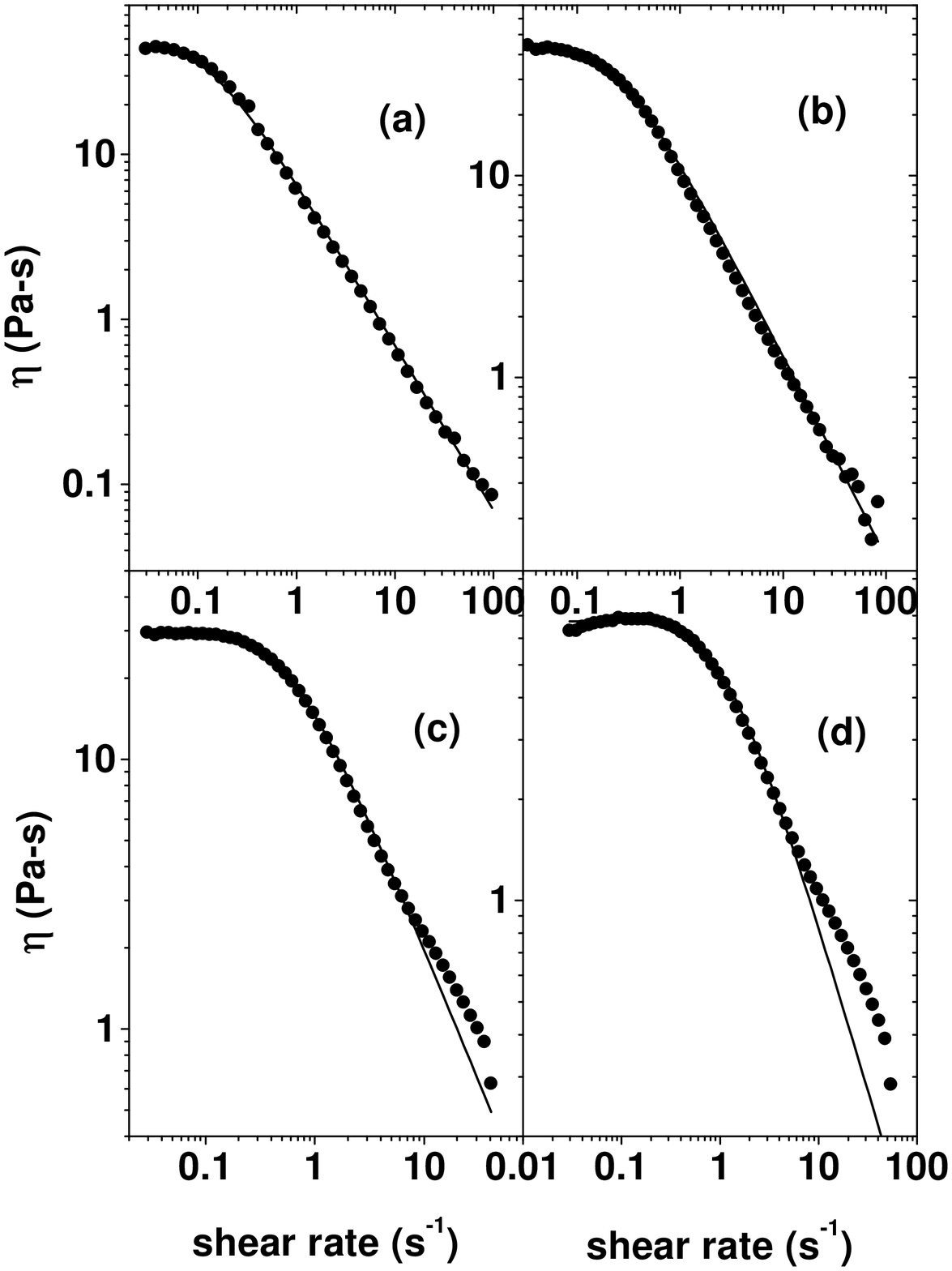}}
\end{figure}
\centerline{Figure 5}

\newpage
\begin{figure}[h]
\centerline{\epsfxsize = 12cm \epsfysize = 12cm \epsfbox{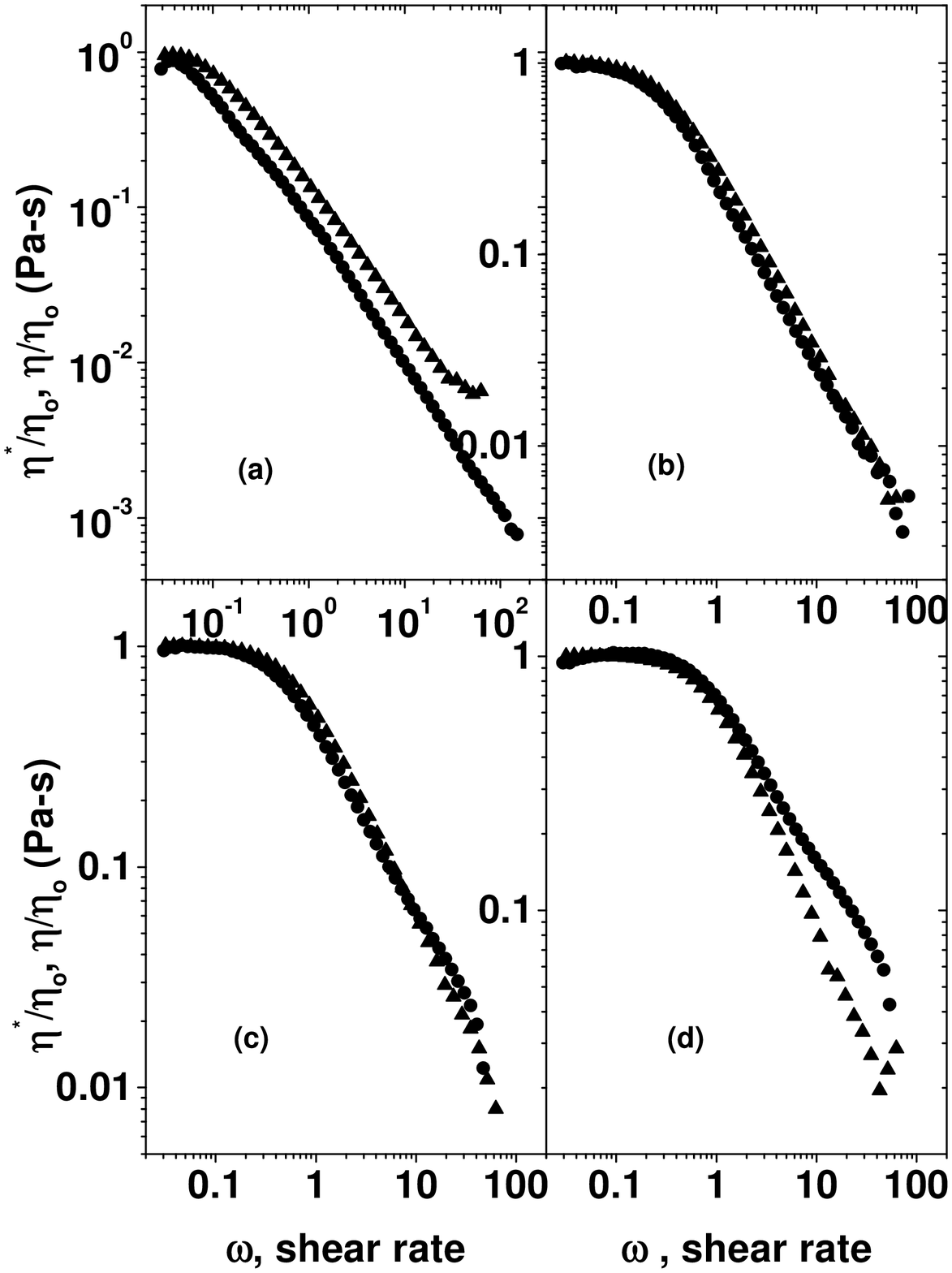}}
\end{figure}
\centerline{Figure 6}

\newpage
\begin{figure}[h]
\centerline{\epsfxsize = 12cm \epsfysize = 12cm \epsfbox{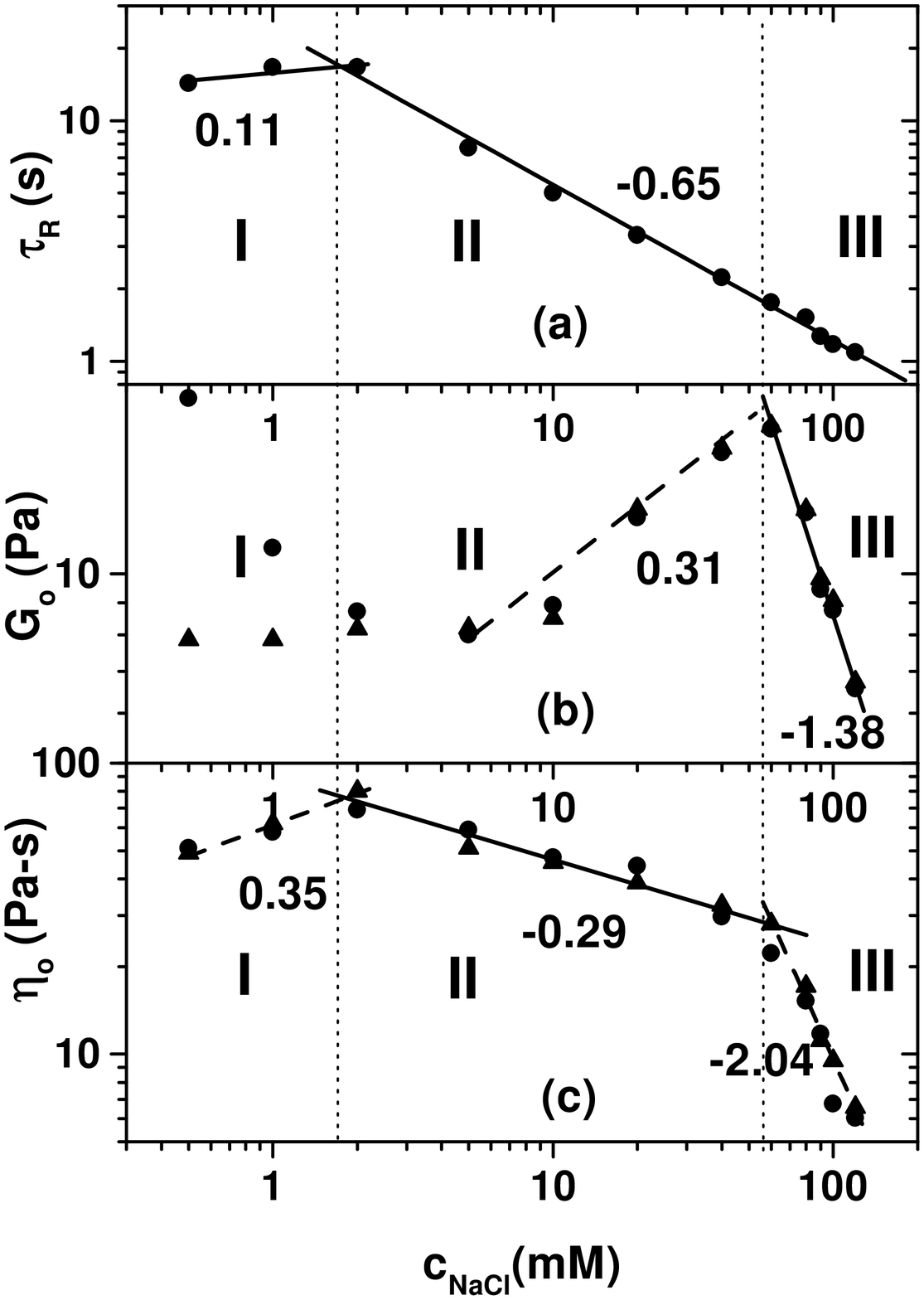}}
\end{figure}
\centerline{Figure 7}

\newpage
\begin{figure}[h]
\centerline{\epsfxsize = 12cm \epsfysize = 12cm \epsfbox{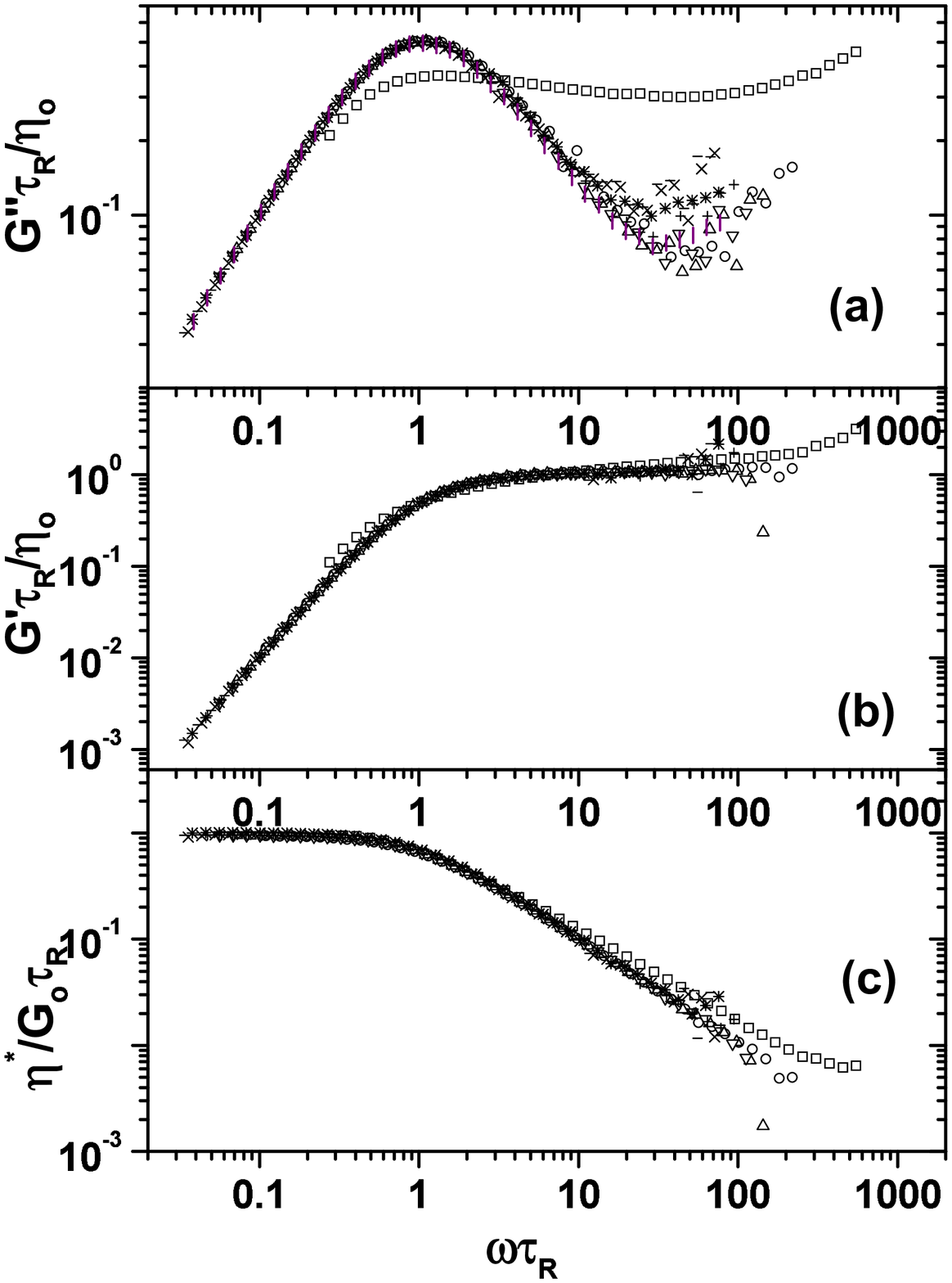}}
\end{figure}
\centerline{Figure 8}

\newpage
\begin{figure}[h]
\centerline{\epsfxsize = 12cm \epsfysize = 12cm \epsfbox{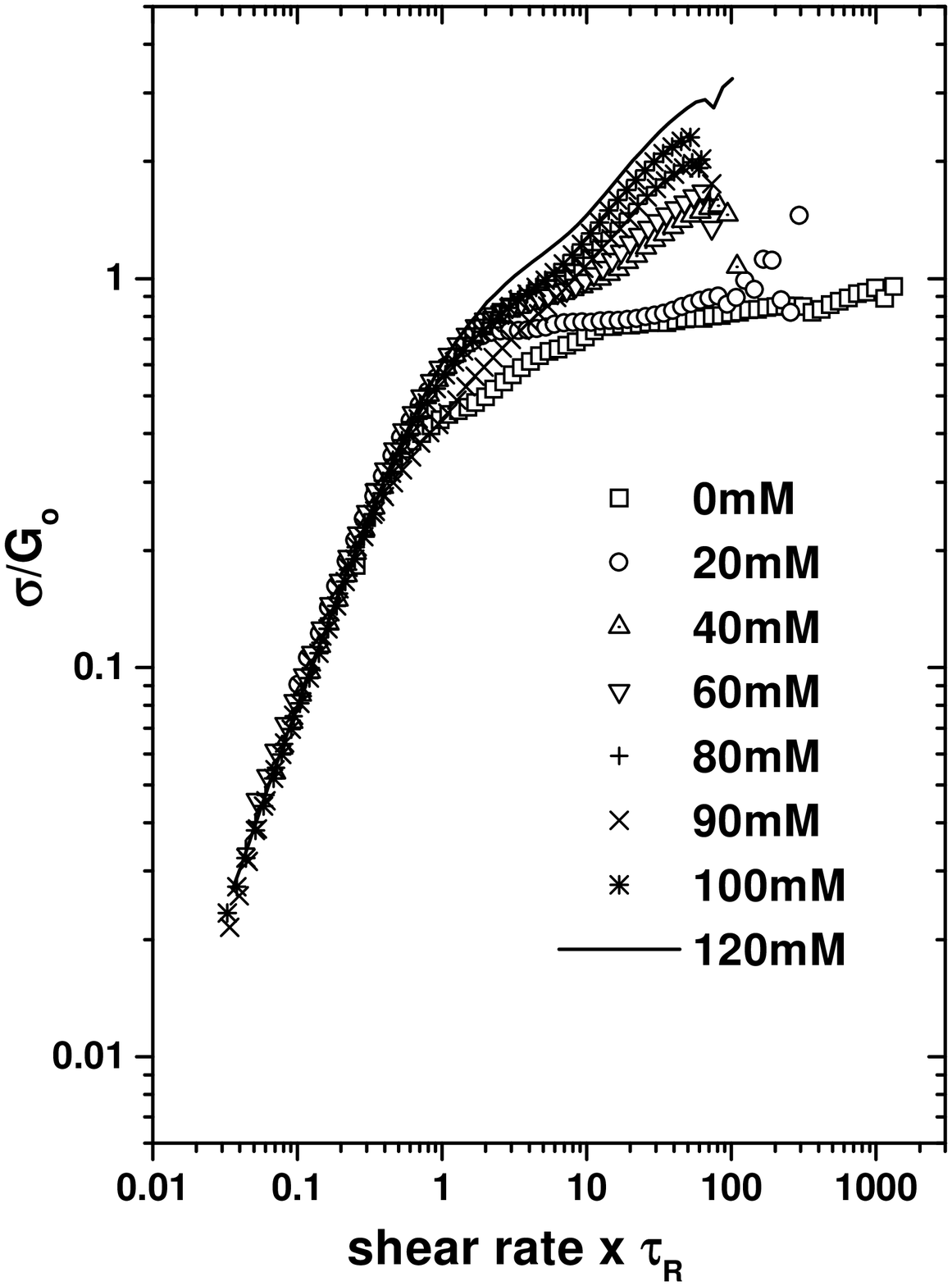}}
\end{figure}
\centerline{Figure 9}

\end{document}